\documentclass[twocolumn,twoside,slac_two]{revtex4}
\usepackage{graphicx}
\usepackage{fancyhdr}
\usepackage{graphics}
\usepackage{epstopdf}
\usepackage{textpos}
\begin{document}

\title{\centering Search for high-mass dilepton resonances with the ATLAS experiment at $\sqrt{s}$ = 7 TeV}


\author{
\centering
\begin{center}
S. Viel
\end{center}}
\affiliation{\centering University of British Columbia - TRIUMF, Canada}
\begin{abstract}
We present a search for high-mass $\ell^{+}\ell^{-}$ resonances in $pp$ collisions at a centre-of-mass energy of 7 TeV recorded by the ATLAS experiment in 2011. No statistically significant excess above the Standard Model expectation is observed in a dataset corresponding to an integrated luminosity of approximately 1 fb$^{-1}$. Consequently, upper limits are set on the cross-section times branching ratio of resonances decaying to muon pairs as a function of the resonance mass.
In particular, a Sequential Standard Model $Z'$ is excluded for masses below 1.83 TeV, and a Randall-Sundrum Kaluza-Klein graviton with coupling $k/\overline{M}_{Pl} = 0.1$ is excluded for masses below 1.63 TeV, both at the 95\%~C.L.
\end{abstract}

\maketitle
\thispagestyle{fancy}



Many hypotheses that go beyond the Standard Model predict the existence of new, high-mass resonances decaying into lepton pairs. We have searched for such resonances~\cite{ref:dilep} using proton-proton collision data at a centre-of-mass energy of 7 TeV, recorded with the ATLAS detector~\cite{ref:ATLAS}.  The dataset corresponds to an integrated luminosity of 1.08 fb$^{-1}$ in the electron channel, and 1.21 fb$^{-1}$ in the muon channel. The benchmark signals considered in this search are new neutral vector bosons ($Z'$) from the Sequential Standard Model (SSM)~\cite{ref:Zp} or models involving an $E_6$ grand unified symmetry group~\cite{ref:E6}, and Randall-Sundrum Kaluza-Klein gravitons ($G^*$)~\cite{ref:Gstar} for couplings $k/\overline{M}_{Pl} \leq 0.1$, where $k$ is the space-time curvature in the extra dimension and $\overline{M}_{Pl} = M_{Pl}/\sqrt{8\pi}$ is the reduced Planck scale.


Three main detector systems of the ATLAS detector are used in this analysis.  The inner detector, used to reconstruct charged particle tracks and vertices, consists of silicon pixels, silicon strips and transition radiation detectors, covering $|\eta| < 2.5$.  It is immersed in a homogeneous solenoidal magnetic field, which allows to measure the momentum of charged particles using track curvature.  The calorimeters, responsible for the reconstruction of particle showers, are made of liquid argon and lead in the electromagnetic part, while the hadronic part is composed of scintillating tiles and iron in the central region and liquid argon, copper and tungsten in the forward region. Outside the calorimeter, toroid magnets provide the field for the muon spectrometer, which consists of resistive-plate and thin-gap trigger chambers, and three sets of drift tubes and cathode strip chambers for high-precision reconstruction of muon tracks.

Electron candidates are identified using clusters in the electromagnetic calorimeter associated with inner detector tracks, while muon candidates are combined tracks from the independent inner detector and muon spectrometer measurements.


This search looks for two opposite-sign muons or electrons, forming a narrow peak in the invariant mass spectrum. 
Lepton pairs are selected in events passing a single-electron trigger requiring a transverse energy $E_T > 20$ GeV, or a single-muon trigger requiring a transverse momentum $p_T > 22$ GeV. The primary vertex of the collision is required to have at least 3 charged particle tracks.

In dielectron events, both electrons are required to have $E_T > 25$ GeV, be within the pseudorapidity range $|\eta| < 2.47$ and avoid the transition region between the barrel and endcap of the electromagnetic calorimeter.  Cuts on the transverse shower shape, leakage into the hadronic calorimeter reduce background from QCD jets, as well as an additional cut on the calorimeter isolation of the leading electron, such that $\Sigma E_T (\Delta R<0.2) < 7$ GeV.  The reconstructed tracks from both electrons are required to satisfy quality requirements and to match clusters in the calorimeter.  Finally, to reject conversion electrons, a hit in the innermost layer of the Pixel detector is required.  The signal acceptance for a 1.5~TeV resonance is 65\% for $Z' \rightarrow e^{+}e^{-}$, and 69\% for $G^{*} \rightarrow e^{+}e^{-}$.

In dimuon events, both muons are required to have $p_T > 25$ GeV, and to pass stringent hit requirements in both the inner detector and muon spectrometer of ATLAS, to ensure that their track momentum is well-measured. This selection includes a three-layer requirement in the muon spectrometer. To suppress background contributions from cosmic rays, the primary collision vertex of the event is required to have at least 3 charged particle tracks, and to be within 20 cm of the centre of the detector. Further, both muon tracks are required to originate within 0.2 mm of the primary vertex in the direction transverse to the beam line, and within 1.0 mm longitudinally. Finally, to reduce background from QCD jets, an isolation requirement is imposed on both muons such that $\Sigma p_T (\Delta R<0.3)/p_T (\mu)<0.05$. The signal acceptance for a 1.5~TeV resonance is 40\% for $Z' \rightarrow \mu^{+}\mu^{-}$, and 44\% for $G^{*} \rightarrow \mu^{+}\mu^{-}$.

\begin{figure}[!h!tpb]
  \centering
  \includegraphics[width=0.49\textwidth]{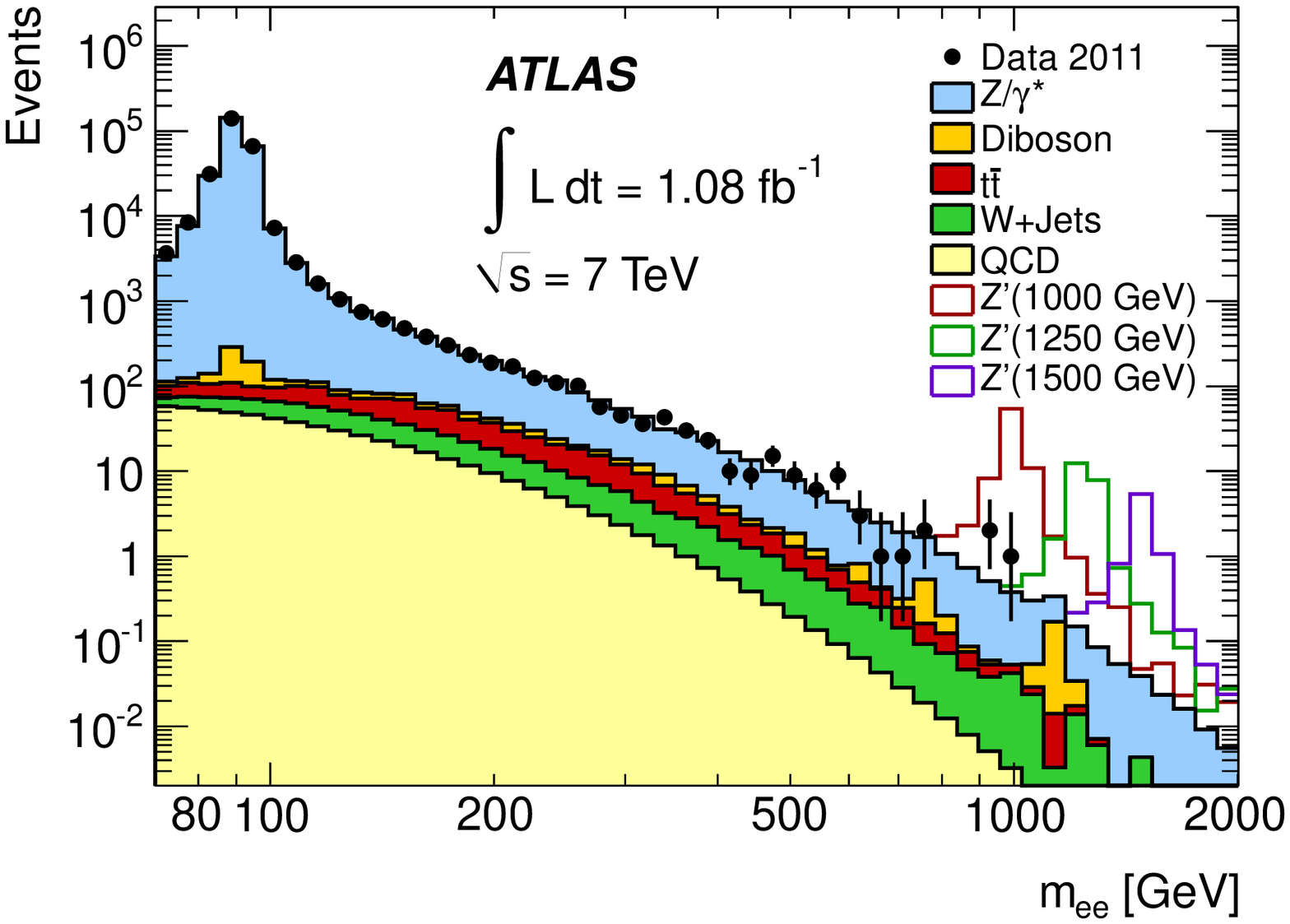}
  \includegraphics[width=0.49\textwidth]{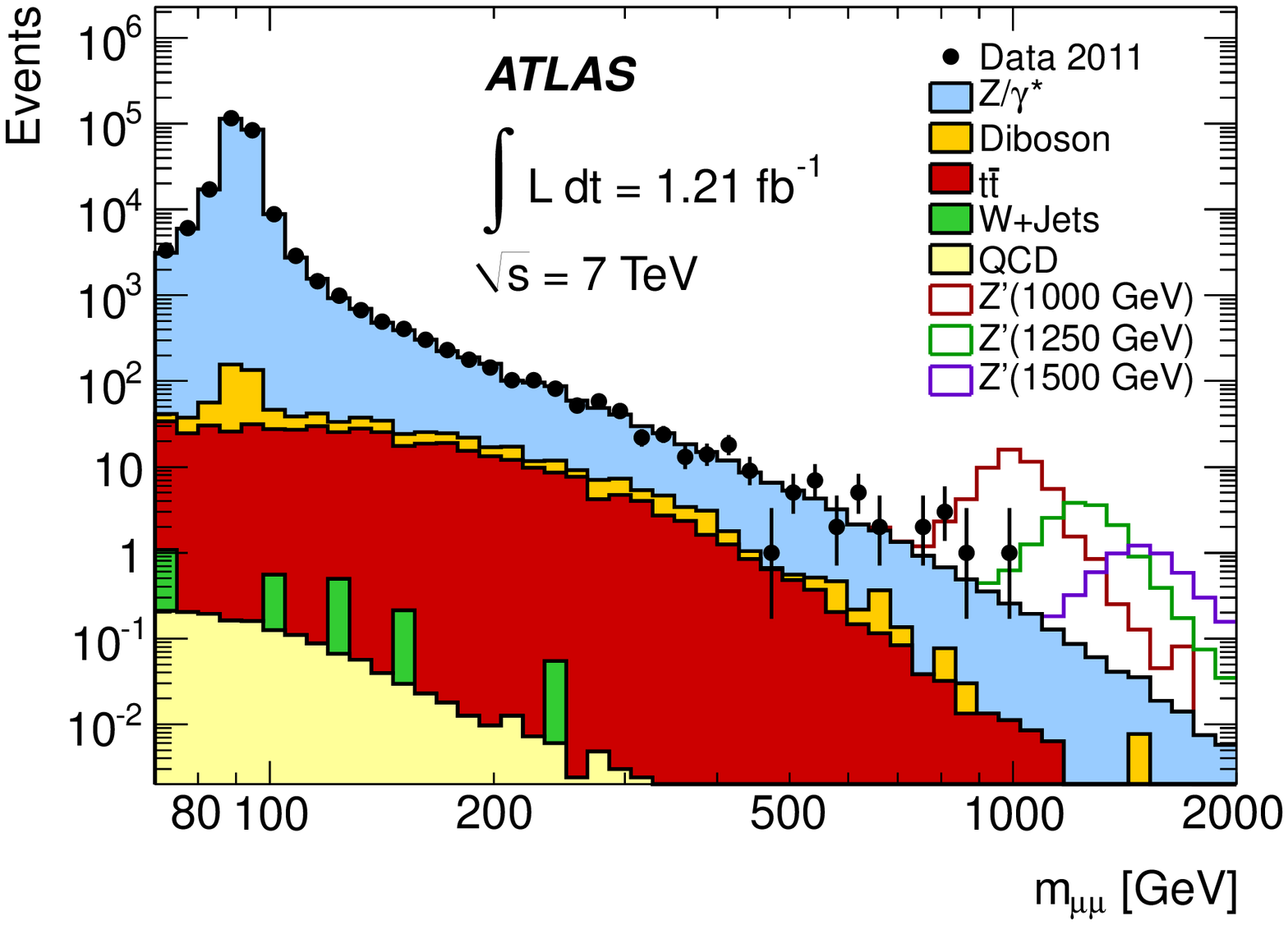}
  \caption{Dielectron (top) and dimuon (bottom) invariant mass distributions after final selection, compared to the stacked sum of all expected backgrounds, with three example $Z'_{SSM}$ signals overlaid. The bin width is constant in log $m_{\ell\ell}$.~\cite{ref:dilep}}
  \label{fig:mass}
\end{figure}

The resulting invariant mass distributions of dielectrons and dimuons are shown on Figure~\ref{fig:mass}. Standard Model backgrounds due to the Drell-Yan process and the production of dibosons, $W$ bosons in association with jets, and top quark pairs are evaluated using Monte Carlo event generators, and the full simulation of the ATLAS detector based on GEANT4.  The background due to QCD multijets is estimated from data.  In the muon channel, a reverse muon isolation cut is used to obtain the QCD shape, which is scaled using the ratio of isolated to anti-isolated dimuons in QCD Monte Carlo.  In the electron channel, the shape is obtained using reverse identification, and a 2-component fit to the data with invariant mass templates from the Monte Carlo and QCD estimates gives their relative normalization.

The total background estimate is scaled to the data in the invariant mass range 70-110 GeV.  For a resonance mass of 1.5 GeV, systematic uncertainties on the signal cross-section due to parton distribution functions and $\alpha_s$ variations amount to 10\%, while uncertainties on the NNLO corrections to the Drell-Yan background are evaluated to 5\%.  The theoretical uncertainty on the $Z$ boson cross-section is taken to be 5\%.  In the muon channel, there are additional uncertainties on the trigger and reconstruction efficiencies, totaling 4.5\%.


To characterize the signal significance of our data, we calculate the probability of observing an excess at least as signal-like as the one observed in data, assuming that signal is absent ($p$-value). The outcome of the search is ranked against pseudo-experiments from background processes using a log-likelihood ratio, scanned as a function of signal cross-section and mass. The resulting $p$-values being 54\% in the electron channel and 24\% in the muon channel, the data are consistent with the Standard Model. 

Given the absence of a signal, an upper limit on the signal cross-section times branching ratio ($\sigma B$) is determined at the 95\% confidence level (C.L.), using a Bayesian approach~\cite{ref:BAT}. The invariant mass distributions from data are compared with the expected background and signal templates. The observed limits are then obtained from a likelihood function, derived from the Poisson probability for the observed number of data events given the expectation from the templates. The median expected limits are obtained from an ensemble of pseudo-experiments using events drawn from the background hypothesis. Finally, mass limits are calculated by comparing the limits on $\sigma B$ to theoretical predictions for different $Z'$ models and $G^*$ couplings.  

\begin{figure*}[!h!tpb]
  \centering
  \includegraphics[width=0.49\textwidth]{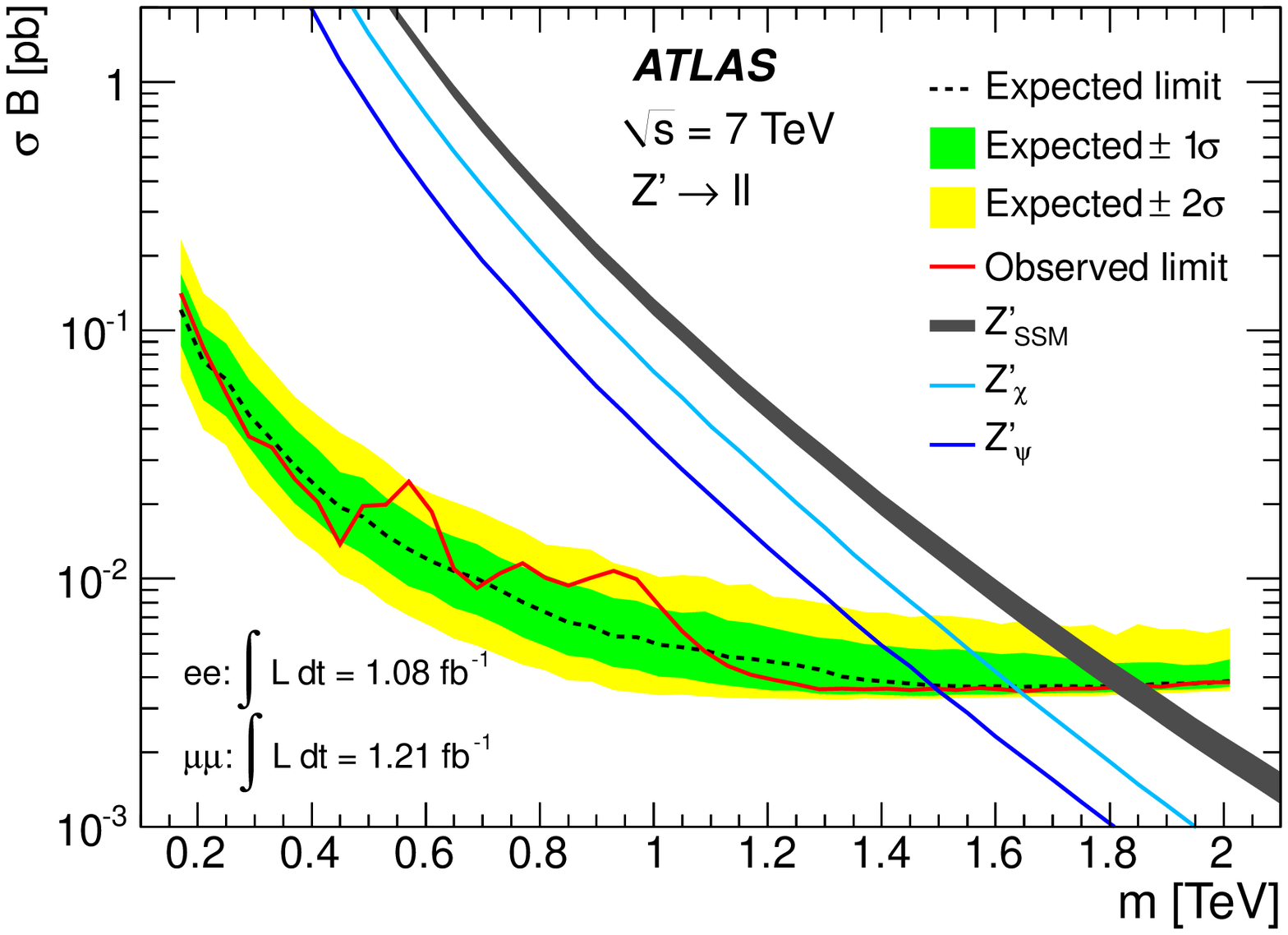}
  \includegraphics[width=0.49\textwidth]{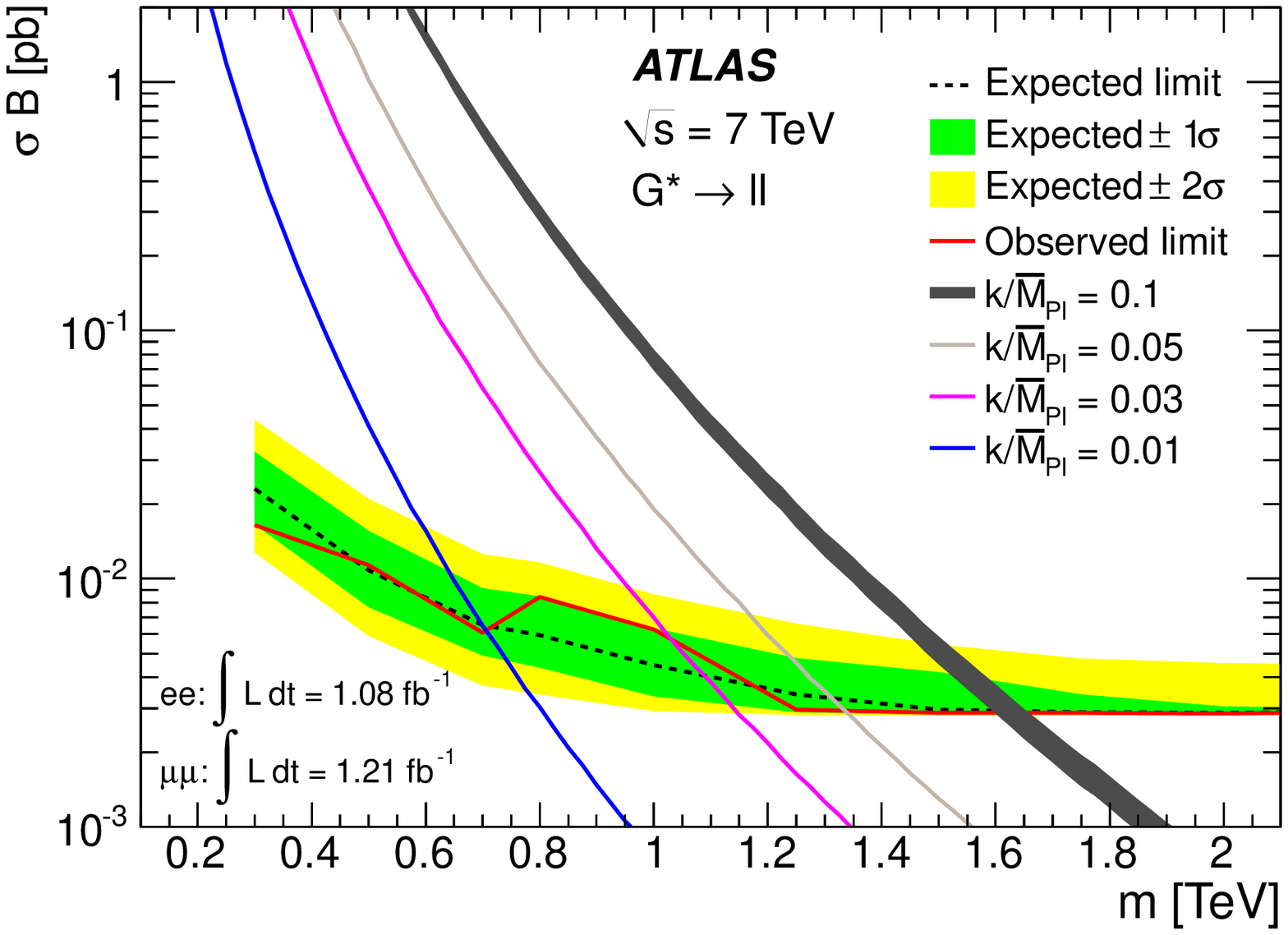}
  \caption{Combined expected and observed 95\%~C.L. upper limits on $\sigma B$ as a function of mass for $Z'$ models (left) and for different $G^*$ couplings (right).~\cite{ref:dilep}}
  \label{fig:lim}
\end{figure*}

\begin{figure}[!h!tpb]
  \centering
  \includegraphics[width=0.49\textwidth]{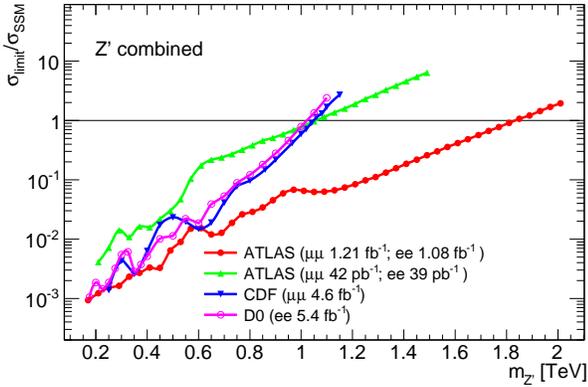}
  \caption{Ratio of the combined observed 95\% C.L. cross-section limit for $Z'$ production, divided by the $Z'_{SSM}$ cross-section.~\cite{ref:dilep}}
  \label{fig:limRatio}
\end{figure}

Combined expected and observed 95\%~C.L. upper limits on $\sigma B$ as a function of mass are shown on Figure~\ref{fig:lim}, along with theoretical curves for $\sigma B$ in the different $Z'$ models and $G^*$ couplings considered. Observed and expected mass limits on $Z'_{SSM}$, and on $G^*$ for $k/\overline{M}_{Pl}$=0.1 are presented in Table~\ref{tab:limits}, while observed limits on $Z'_{E_6}$ models and on $G^*$ for different values of $k/\overline{M}_{Pl}$ are given in Table~\ref{tab:limits2}. 
Finally, Figure~\ref{fig:limRatio} shows the ratio of the combined observed 95\%~C.L. cross-section limit for $Z'$ production, divided by the $Z'_{SSM}$ cross-section.

\begin{table}[!htb]
\caption{Observed (Expected) 95\%~C.L. mass lower limits in TeV on $Z'_{SSM}$ and $G^*$ resonances.
}
\label{tab:limits}
\begin{center}
\resizebox{\columnwidth}{!}{
\small
\begin{tabular}{l c c c}
\hline
Model~~~~~          &      $ ee$~~~      &       $\mu\mu$~~~  &   $\ell\ell$~~   \\
\hline			 	 	 	 	  	 
$Z'_{SSM}$            &   1.70~(1.70)  &    1.61~(1.61) &   1.83~(1.83) \\
$G^* $  ($k/\overline{M}_{Pl}=0.1$)   &   1.51~(1.50)     &    1.45~(1.44) &	1.63~(1.63)      \\
\hline
\end{tabular}
}
\end{center}
\end{table}

\begin{table}[!htb]
\caption{Combined observed 95\%~C.L. mass lower limits in TeV on $Z'_{E_6}$ models and on $G^*$ for different values of the coupling $k/\overline{M}_{Pl}$. 
}
\label{tab:limits2}
\begin{center}
\resizebox{\columnwidth}{!}{
\begin{tabular}{l|cccccc|cccc}
\hline
& \multicolumn{6}{c|}{$Z'_{E_6}$ Models}   & \multicolumn{4}{c}{$G^* $} \\
\hline
Model/Coupling            & $Z'_{\psi}$ & $Z'_{N}$  & $Z'_{\eta}$ & $Z'_{I}$  & $Z'_{S}$ & $Z'_{\chi}$ & 0.01 & 0.03  & 0.05 & 0.1 \\
\hline			 	 	 	 	  	 
Mass limit [TeV] & 1.49  & 1.52  & 1.54  & 1.56 & 1.60 & 1.64 & 0.71 & 1.03  & 1.33  & 1.63  \\
\hline
\end{tabular}
}
\end{center}
\end{table}

In conclusion, using over 1 fb$^{-1}$ of proton-proton data recorded by ATLAS, we have searched for narrow dilepton resonances in the invariant mass spectrum. Observations are consistent with Standard Model expectations. We therefore set 95\%~C.L. limits on various $Z'$ models and on Randall-Sundrum Kaluza-Klein gravitons.

Future goals of this analysis include increasing the signal acceptance in both channels and setting limits on a wider range of theoretical models.

\bigskip 
\begin{acknowledgments}
The author acknowledges support from the Vanier Canada Graduate Scholarship program, and the Natural Sciences and Engineering Research Council of Canada. 
\end{acknowledgments}

\bigskip 
\bibliography{basename of .bib file}

\end{document}